\documentclass[aps,preprint]{revtex4}
\usepackage{amsmath}
\usepackage{epsfig}
\usepackage{float}
\usepackage{dcolumn} 

\DeclareMathAlphabet{\mathpzc}{OT1}{pzc}{m}{it}
\newcommand{\be}{\begin{equation}}
\newcommand{\ee}{\end{equation}}
\newcommand{\bea}{\begin{eqnarray}}
\newcommand{\eea}{\end{eqnarray}}
\newcommand{\beas}{\begin{eqnarray*}}
\newcommand{\eeas}{\end{eqnarray*}}

\begin{document}
\title{\bf Supersymmetric partner potentials arising from nodeless half bound states}  
\author{Zafar Ahmed$^1$, Dhruv Sharma$^2$, Rahul Kaiwart$^3$, Mohammad Irfan$^4$}
\affiliation{$~^1$Nuclear Physics Division, Bhabha Atomic Research Centre, Mumbai 400 085, India \\
$~^2$Department of Physics, National Institute of Technology, Rourkela, 769008, India\\
$~^3$Human Resource Development Division, Bhabha Atomic Research Centre, Mumbai, 400 085, India\\
$~^4$Department of Physics, Indian Institute of Science Education and Research, Bhopal,  462066 India}
\email{1:zahmed@barc.gov.in, 2:sharmadhru.gmail.com, 3: rahul.kaiwart@gmail.com, mohai@iiserb.ac.in}
\date{\today}
\begin{abstract}
\noindent
A Half Bound State (HBS) $\psi_*(x)$ can be defined as a single, conditional, zero-energy, continuous  solution of the one dimensional Schr{\"o}dinger equation for a  scattering potential well $V(x)$ ($s.t ~ V(\pm \infty)=0$). The non-normalizable and solitary HBS of a potential satisfies Neumann boundary condition that $\psi'_*(\pm \infty)=0$ and it can have  $n$ (= 0,1,2,...) number of nodes indicating $n$ number of bound states in $V(x)$ below $E=0$. Here we show that starting with a nodeless HBS, we can construct a (supersymmetric)  pair of finite potentials (well, double wells, well-barrier): $V_{\pm}(x)$  having no bound state and they enclose positive area on $x$-axis. On the contrary their negative counterparts $(-cV_{\pm}(x)),c>0$ do have at least one bound state for any arbitrary positive value of $c$. Furthermore, $c V_{\pm}(x),~ c >0$ which binds positive area on x-axis in conformity with Simon's theorem can have at least one bound state only conditionally for instance when $c>1$ or $c>>1$.
\end{abstract}
\maketitle
The concept of supersymmetry is well known to have provided a re-formulation of quantum mechanics in terms of solution of Schr{\"o}dinger equation for a potential $V(x)$. Eventually, we have  the well developed [1-3] supersymmetric quantum mechanics giving rise to
new techniques and some new solvable models. The best known of these results is the creation of supersymmetric partner potentials $V_{\pm}(x)$
\begin{equation}
V_{\pm}(x)=W^2(x) \pm W'(x), \quad  W(x)=-\frac{\psi'_0(x)}{\psi_0(x)}
\end{equation}	
starting with a normalizable ground state $\psi_0(x)$. These two potentials have identical spectrum excepting that $V_+(x)$ misses the ground state eigenvalue of $V_-(x)$. Interestingly, if a potential well $V_-(x)$ has just one bound state, consequently $V_+(x)$ becomes a barrier possessing only scattering states.

Here in this note, we suggest the use of a nodeless zero energy Half Bound State (HBS: $\psi_*(x)$) in place of $\psi_0(x)$. We  show that the potentials $V_{\pm}(x)$ hence, constructed are wells, double wells and well-barrier systems which enclose positive area on x-axis and they possess no bound state. However, their negative counter parts $(-V_\pm(x))$ have at least one bound state. We believe that this extends the scope of studying more interesting potential wells, double wells and well-barrier systems. So, here are a variety of one dimensional potentials which have none
or at least one bound state, bringing Simon's Theorem (Theorem 2.5 in [4]) for the potentials $V(x)$ of the type
\begin{equation}
\int_{-\infty}^{\infty} (1+x^2)|V(x)| dx <\infty,
\end{equation}
in to contention.

Half Bound State (HBS) has been discussed in a meager way in the formation of deuteron (neutron-proton) [5] and in low energy scattering of neutron and proton in terms of scattering length [6]. The concept of HBS  could be seen as an amusing and instructive feature of attractive scattering potential wells in one dimension. By attractive scattering wells we mean the finite potentials $V(x)$ which are either of finite support ($V(|x|>a)=0$) or which vanish asymptotically: $V(\pm \infty)=0$. Recently, while studying the paradoxical zero reflection at zero energy $R(0)=0$ [7], we find that if a scattering well has a HBS at $E=0$, it is then $R(E) \rightarrow 0$ as
$E\rightarrow 0$. We denoted [8] HBS as $\psi_*(x)$ and found that as a single, conditional, zero-energy and continuous  solution of the one dimensional Schr{\"o}dinger equation for $V(x)$. The HBS satisfies Neumann boundary condition that $\psi'_*(\pm \infty)=0$ and it can have   $n(= 0,1,2,3...)$ number of node(s) indicating $n$ number of bound states in $V(x)$ below $E=0$. Using the simple exponential and square wells we have demonstrated HBS  at $E=0$ of one or more than one nodes as the depth of the well is increased. In simple single wells we remarked that the nodeless HBS does not exist or it is just a constant i.e., $\psi_*(x)=C$. In this paper, we suggest the possibility of a non-constant nodeless HBS as 
\begin{equation}
\psi_*(x)={\cal A}+{\cal F}(x) \ne 0 \quad \forall \quad x \in (-\infty, \infty), \quad {\cal F}(\pm \infty)=C_1,C_2, \quad {\cal A} \in {\cal R}.
\end{equation}
Here ${\cal F}(x)$ is a symmetric or an asymmetric continuous function of $x$ which could also be a physical ground state or excited eigenstate of a one dimensional potential.
\begin{figure}[t]
\centering
\includegraphics[width=5 cm,height=5.cm]{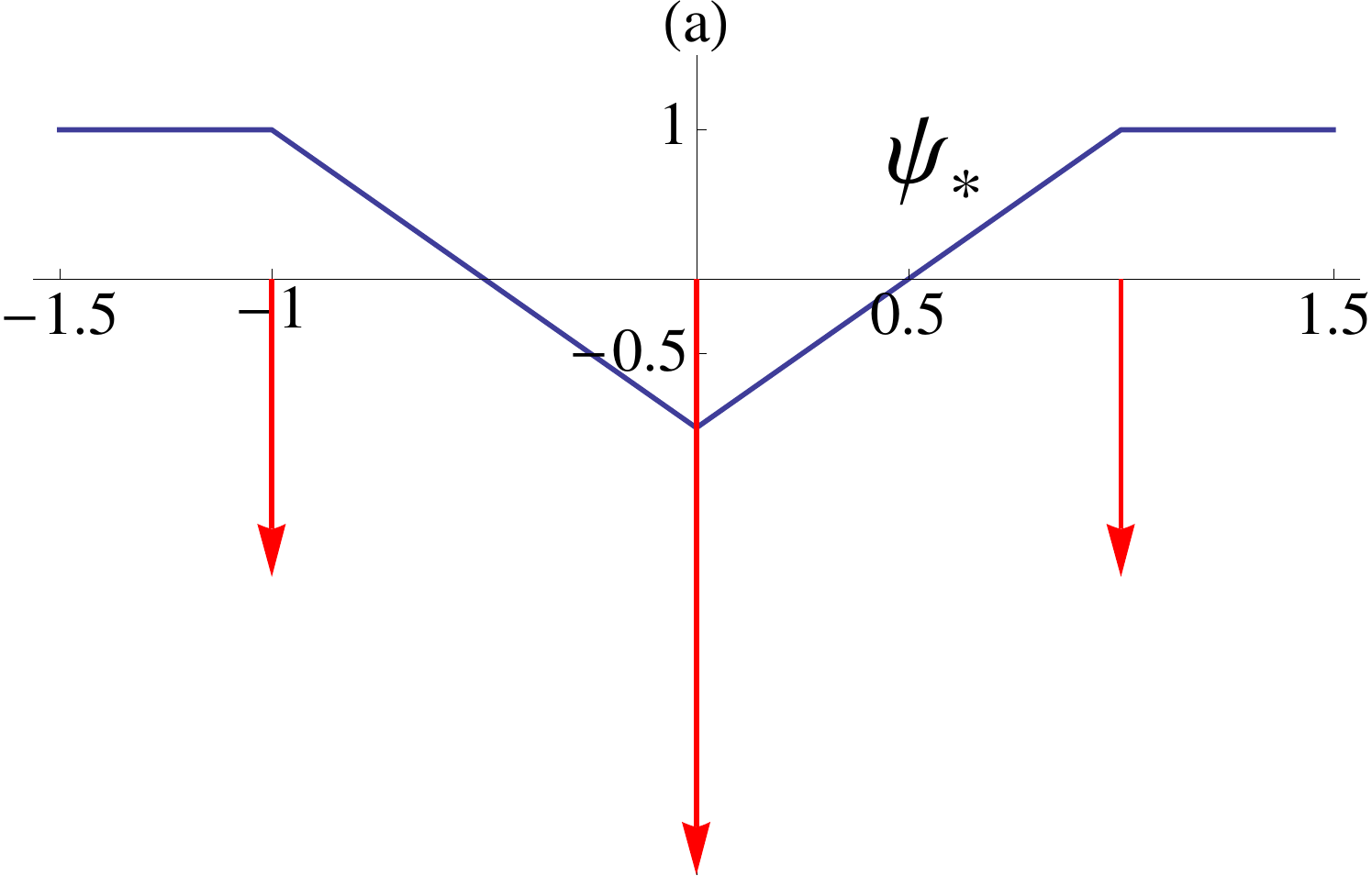}
\hspace{.25 cm}
\includegraphics[width=5 cm,height=5.cm]{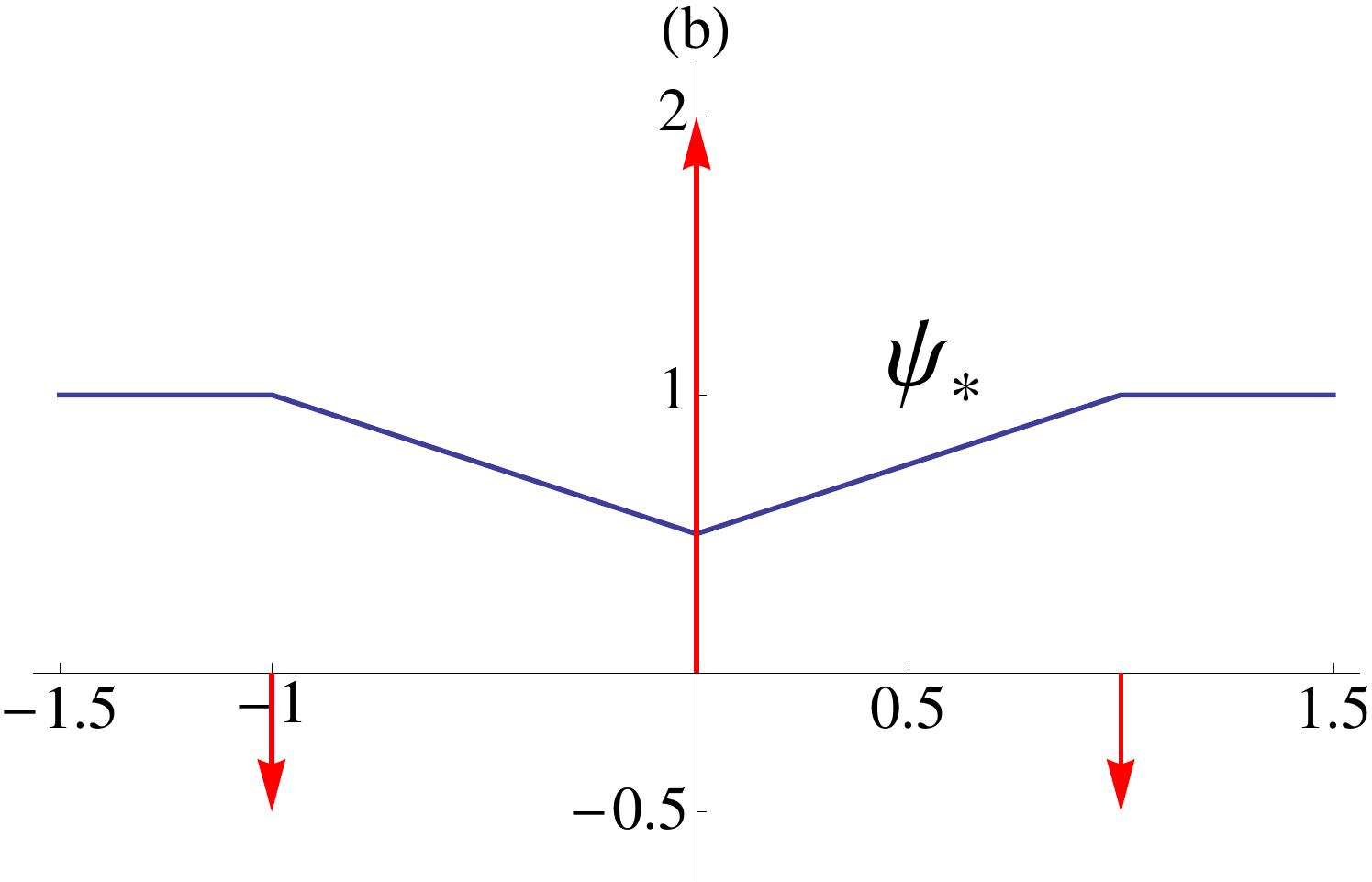}
\hspace{.25 cm}
\includegraphics[width=5. cm,height=5.cm]{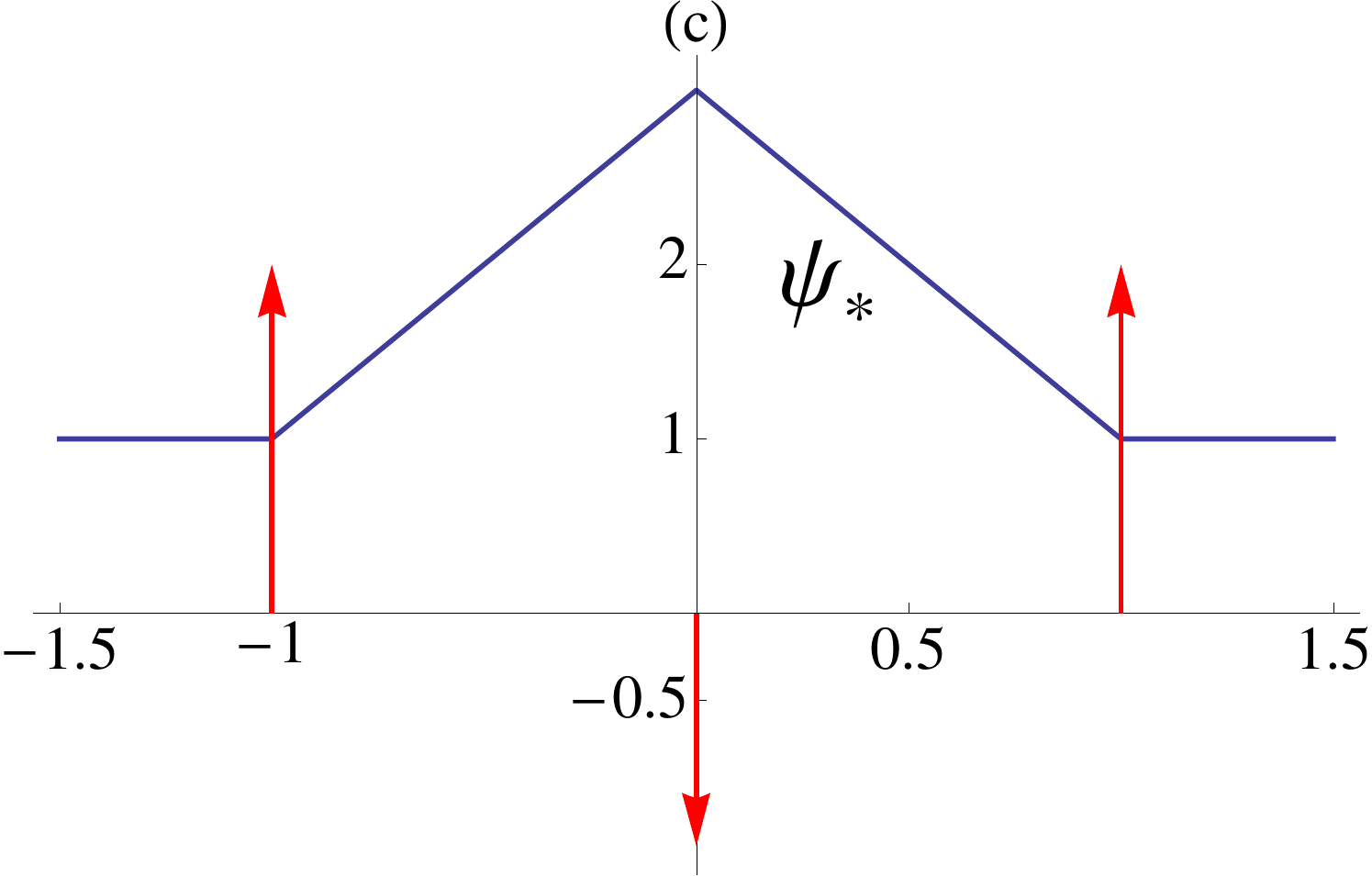}
\caption{Depiction of HBS in triple Dirac delta potential well (3) (a) $U_1=2, U_2=4$, (b) $U_1=1/2, U_2=-2$, (c) $U_1=-2, U_2=4/3$. In (a), HBS has two nodes indicating two bound states at $E<0$ in the triple delta well. Parts (b,c) depict non-constant nodeless HBS indicating no bound state below $E=0$. The area integrals $I$ (8) in (a) is negative and in (b,c) it is positive. The upward (downward) arrows indicate repulsive (attractive) Dirac delta potentials.}
\end{figure} 

First, in the following we would like to present a triple Dirac delta potential as a simple solvable model of non-constant nodeless HBS. We construct the triple well Dirac delta potential as
\begin{equation}
V(x)=-U_1\delta(x+a)-U_2\delta(x)-U_1\delta(x-a),
\end{equation}
to be solved in the Schr{\"o}dinger equation 
\begin{equation}
\frac{d^2\psi(x)}{dx^2}+[E-V(x)]\psi(x)=0,
\end{equation}
here again we have set $2\mu=1=\hbar^2$ and $U_j>0$ means delta potential is a well.
For zero energy the solution of this potential (4) could be only linear $\psi_*(x)= \alpha x+ \beta $ with zero-curvature [9]. So for $E=0$, we write 
\begin{equation}
\psi_*(x)= \left\{\begin{array}{lcr}
A, & & x<-a ,\\
Bx+C, & & -a\le x < 0,\\
Dx+F, & & 0\le x < a,\\
A, && x \ge a
\end{array}
\right.
\end{equation}
By demanding the continuity of $\psi_*(x)$ and discontinuity [11] of its first derivative (due to Dirac Delta functions) at $x=-a,0,a$, we get six equations and solving them we get
\begin{equation}
B=-U_1 A, \quad C=(1-U_1a) A=F, \quad D=-(U_1+U_2-U_1U_2a) A, \quad U_2=2U_1/(U_1a-1).
\end{equation}
Three cases arise here (i) $U_1a>1$, we have triple Dirac Delta wells and $\psi_*(x)$ is a two node state HBS indicating possibility of two bound states for $E<0$ (see Fig. 1(a)). (ii) $0<U_1 a<1$ making $U_2 a<0$ namely two delta wells at $\pm a$ and a delta barrier at $x=0$, then $\psi_*(x)$ is a nodeless HBS (Fig. 1(b)). (iii) $U_1 a<0$, namely two delta barriers at $x=\pm a$ and one delta well at $x=0$, we again get a nodeless HBS (Fig. 1(c)). 

The nodeless states depicted in Fig. 1(b,c) are piecewise zero curvature HBS. Next, we propose to use the ansatz (2) as  nodeless HBSs which are  differentiable everywhere, as these are a constant ${\cal A}$ added to a physical eigenstate $\psi_0(x).$ Now, we choose $\psi_*(x)={\cal A}+ e^{-x^2}$ to insert in Eq. (1), to construct $V_\pm(x)$  for ${\cal A}=1/2,1,-2$, these potential are depicted as solid and dashed lines respectively in Fig. 2. Notice that these are wells or double wells as they bind a  nodeless HBS at $E=0$, they cannot have a bound state below energy $E=0$. 

Thus, the constructed potentials $V_{\pm}(x)$ have no  bound state. However, their negative partners $-V_{\pm}(x)$ as per Simon's criterion [4] will have at least one bound state. The presence or the absence  bound states can be easily verified by Griffith's  ``wag-the-dog" method: a  one line {\it Mathematica} program [10]. We find that the ground state eigenvalues of $-V_-(x)$ for three cases (${\cal A}=1/2,1,-2$) occur at $E=-0.2432, -0.07344$ and $-0.3127$, respectively. For $-V_+(x)$  the ground state eigenvalues are at $E=-0.5837, -0.2151$ and $-0.0924$, respectively.

A well surrounded by two side barriers such that there are four real turning points at positive energy is one of the {\it necessary} (but not {\it sufficient}) condition for a potential to possess shape resonances [11]. These resonances appear as well separated thin peaks $(T(E_n)=1)$. Using the method [9] based on the numerical integration of Schr{\"o}dinger equation for the scattering potentials $V_{\pm}(x)$ discussed here, one can calculate the transmission $T(E)$ and reflection $R(E)$ coefficients.  In these cases (Fig. 2), we do not find  a sharp peak in  $T(E)(=1)$, however, $R(E)$ becomes very small at higher energies: $R(E= 4.7)=0.25 \times 10^{-3}, R(E=5.2)=0.16 \times 10^{-3}$ and $R(E=15.2)=0.16 \times 10^{-5}$, respectively in these cases  (Fig. 2).

In Fig. 2, we present $V_{\pm}(x)$ arising from an asymmetric HBS  (a): $\psi_*(x)=2+\tanh x$, (b): $\psi_*(x)= 2+\mbox{erf}(x)$ and (c): $\psi_*(x)= 2+x e^{-x^2}$, respectively. For these asymmetric HBS from Eq. (1), we get well-barrier systems (Fig. 3) which are devoid of a bound state.  For $\psi_*(x)={\cal A}+ \tanh x$, we get 
\begin{equation}
V_+(x) = \frac{2\mbox{sech}^2 x(1+ {\cal A} \tanh x)}{({\cal A}+\tanh x)^2},~  V_-(x) = \frac{-2\mbox{sech}^2 x \tanh x}{({\cal A}+\tanh x)}, ~ g = \frac{1}{2}\log \left(\frac{{\cal A}-1}{{\cal A}+1}\right),
\end{equation}
Where, $V_-(0) = 0$ and $V_+(g) = 0$, The well-barrier systems $V_{\pm}(x)$ in Fig. 3, are roughly mirror image of each other about $x$ equal to some constant. However, curiously for this special case of (8) it follows exactly, that $V_+(g-x) = V_-(x)$.

The {\it necessary and sufficient} condition  (see Theorem 2.5 in [4]) for a potential $\lambda V(x), \lambda>0$ of the type (2) to have at least one bound state is that it should enclose negative area  on $x$-axis:

\begin{figure}[t]
	\centering
	\includegraphics[width=5 cm,height=5.cm]{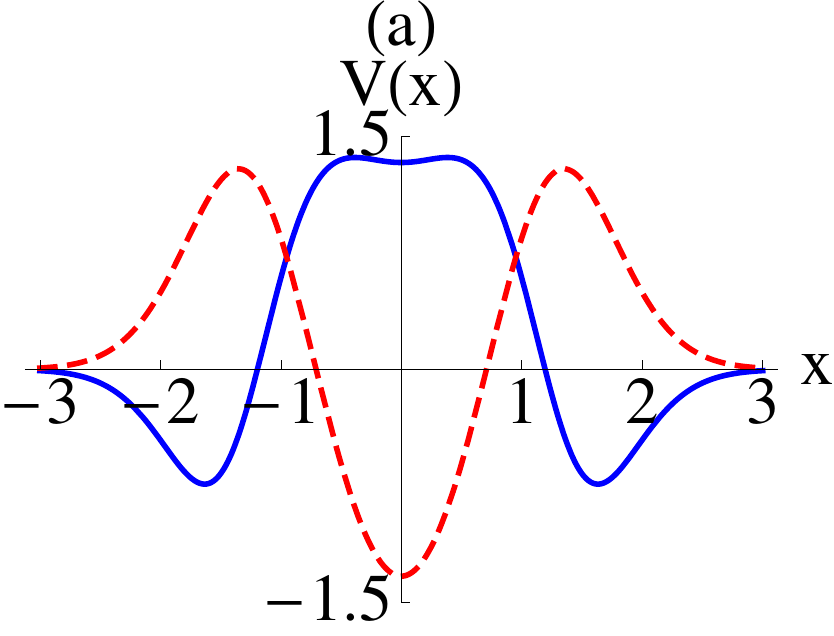}
	\hspace{.25 cm}
	\includegraphics[width=5 cm,height=5.cm]{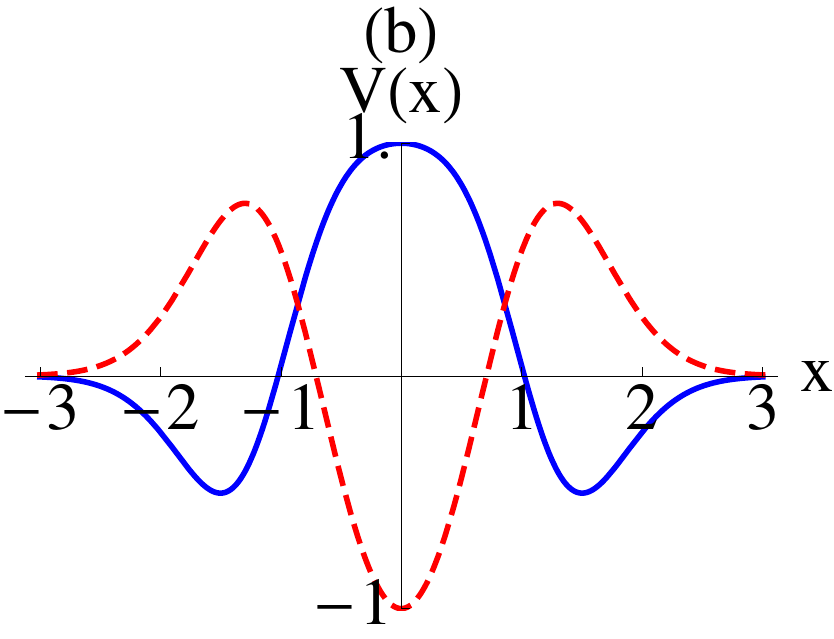}
	\hspace{.25 cm}
	\includegraphics[width=5 cm,height=5.cm]{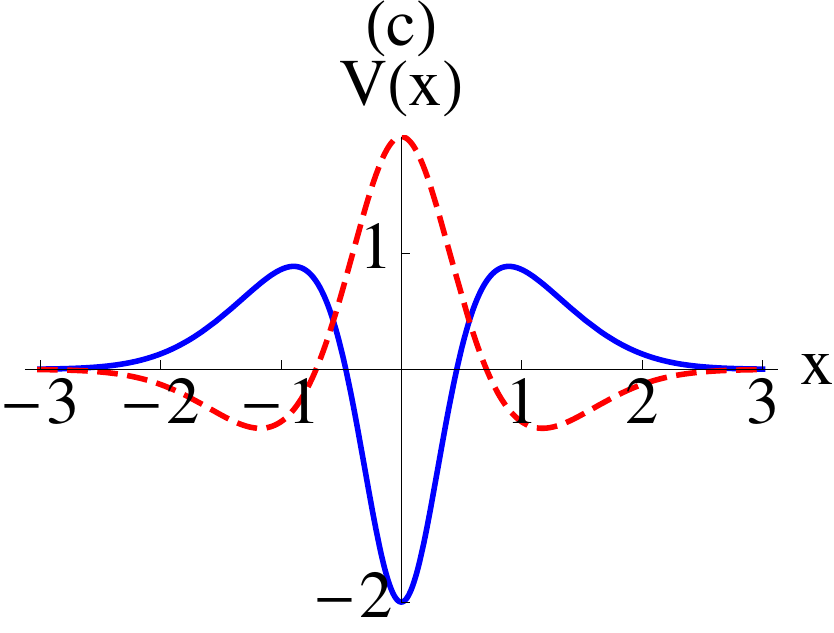}
	\caption{Depiction of $V_-(x)$ (dashed),$V_+(x)$ (solid) arising from Eq. (2) by using the nodeless HBS $\psi_*(x)={\cal A}+ e^{-x^2}$, for ${\cal A}=1/2,1,-2$ respectively in a,b,c. These are wells and double wells yet devoid of eigenvalues as they can bind a nodeless HBS at $E=0$ and hence no bound state for $E<0$. These pairs of potentials enclose positive area 1.38, 0.56, 0.64 , in (a,b,c), respectively are devoid of a bound state also in conformity with the criterion of Simon  for no bound state.}
\end{figure}

\begin{figure}[t]
	\centering
	\includegraphics[width=5 cm,height=5.cm]{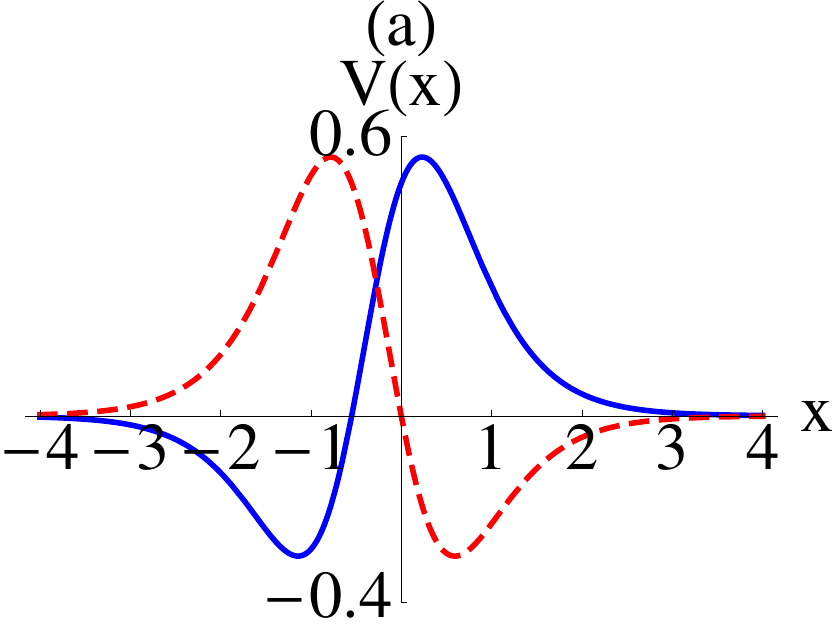}
	\hspace{.25 cm}
	\includegraphics[width=5 cm,height=5.cm]{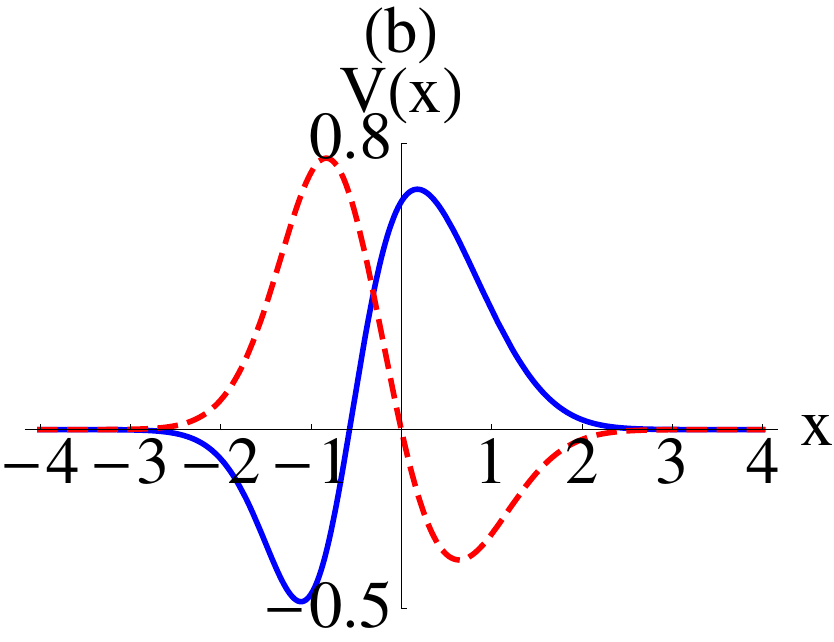}
	\hspace{.25 cm}
	\includegraphics[width=5 cm,height=5.cm]{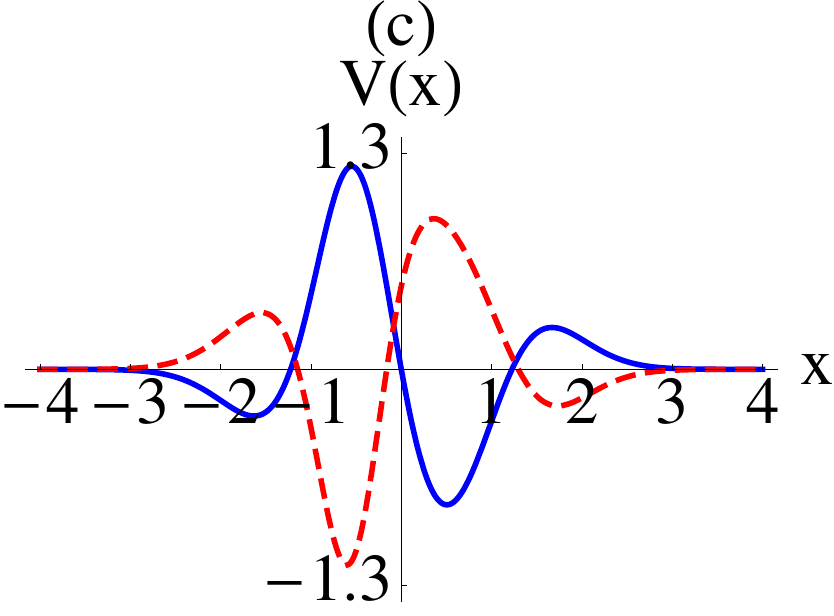}
	\caption{The same as Fig. 2 for (a): $\psi_*(x)={2+ \tanh x}$, (b): $\psi_*(x)={2+ \mbox{erf} x}$, (c): $\psi_*(x)={2+ xe^{-x^2}}$. Notice that these are well-barrier systems.}
\end{figure}

\begin{equation}
I = \int_{-\infty}^{\infty} V(x) dx < 0.
\end{equation}
Here we show that the supersymmetric partners $V_{\pm}(x)$ created by the HBS (3) which by design are devoid of a bound state, have the area integral $I=$ 
\begin{eqnarray}
\int_{-\infty}^{\infty} V_-(x) dx &=& \int_{-\infty}^{\infty} W^2(x) dx- W(\infty)+W(-\infty),\nonumber  \\ 
\int_{-\infty}^{\infty} V_+(x) dx &=& \int_{-\infty}^{\infty} W^2(x) dx + W(\infty)-W(-\infty), \quad W(\pm \infty)=0. 
\end{eqnarray}
as positive definite. On the contrary, the potentials $-V_{\pm}(x)$ enclose negative area on $x$-axis ($I<0$) and we find that they have at least one bound state. This implies that
nonpositivity of $I$ (10) is at least {\it sufficient} for at least one bound state in one dimension.

Further, the magnified potentials: $c V_{\pm}(x)$ ($V_{\pm}(x)$ in Figs. 2-3 multiplied by $c>>1$)  have $I>0$. One or more number of bound states are not unexpected in them, as we get deeper wells (see Figs. 2-3). We find that it is indeed true, for instance for $1.1 V_+(x)$, $I =$  1.52112, we get $E_0=-0.00063$ in the double wells (see solid curve in Fig. 2(a)). The ground state eigen value for  $1.1 V_-(x)$ (see dashed curve in Fig. 2(a)) is $E_0=-0.01990$.   Similarly, in other cases as in Figs. 2-3, we find that the designed poptentials: $V_{\pm}(x)$  using a HBS have $I>0$; they are of the type (2) [4]  and they are devoid of any eigenvalue. But, the scaled ones ($cV_{\pm}(x)$) do have an eigenvalue conditionally ($c>1$) as they the area $I>0$ (10).

We therefore conclude that the unconditional existence  of at least one bound state in the one-dimensional potentials discussed here is as per the Theorem (2.5 in [4]) of Simon.
 
We hope that our suggestion of nodeless half bound state and its role in creating supersymmetric partner potentials (wells, double wells and well-barrier) not having a bound state, will be found interesting and instructive. It is not surprising that such potentials will generate further interest in theory and experiments in future.

\end{document}